\documentclass[twocolumn,aps,prl,showpacs,raggedbottom,nobalancelastpage,amssymb]{revtex4-1}

\usepackage{graphicx}
\usepackage{amsmath}
\usepackage{amssymb}
\usepackage{bm}
\usepackage[usenames]{color}
\usepackage{amsfonts}
\usepackage{appendix}
\usepackage{dsfont}

\def \etal{{\it et al.}~}

\def \Vod{V^{\text{od}}}
\def \Vd{V^{\text{d}}}
\def \VH{V^{\text{H}}}
\def \VF{V^{\text{F}}}
\def \VS{V^{\text{S}}}
\def \lod{\lambda^{\text{od}}}
\def \ld{\lambda^{\text{d}}}

\def \H{\mathcal{H}}
\def \dH{_\text{d}^\text{H}}
\def \oH{_\text{od}^\text{H}}
\def \vSl{v^\text{S}_l}
\def \d{\text{d}}
\def \w{\omega}

\begin{document}
\title{Topological Equivalence between the Fibonacci Quasicrystal and the Harper Model}
\author{Yaacov E.~Kraus and Oded Zilberberg}
\affiliation{Department of Condensed Matter Physics, Weizmann Institute of Science, Rehovot 76100, Israel}


\begin{abstract}
One-dimensional quasiperiodic systems, such as the Harper model and the Fibonacci quasicrystal, have long been the focus of
extensive theoretical and experimental research. Recently, the Harper model was found to be topologically nontrivial. Here, we
derive a general model that embodies a continuous deformation between these seemingly unrelated models. We show that this
deformation does not close any bulk gaps, and thus prove that these models are in fact topologically equivalent. Remarkably, they
are equivalent regardless of whether the quasiperiodicity appears as an on-site or hopping modulation. This proves that these
different models share the same boundary phenomena and explains past measurements. We generalize this equivalence to any
Fibonacci-like quasicrystal, i.e.~a cut and project in any irrational angle.
\end{abstract}


\maketitle

Recent experimental developments in photonic crystals \cite{Negro03,Yoav09} and ultracold atoms \cite{Roati08,Chabe08,Modugno10}
have made the study of the dynamics of particles in one-dimensional (1D) quasiperiodic systems experimentally accessible. These
fascinating systems have long been the focus of extensive research. They have been studied mainly in the context of their
transport and localization properties, showing a variety of interesting transitions between metallic, localized, and critical
phases~\cite{AA,Kohmoto:1983,Han:1994,Hiramoto:1992}. With their recently found nontrivial connection to topological phases of
matter~\cite{us} there is a growing interest in their boundary phenomena~\cite{citing_us1,citing_us2}.

The behavior of particles in such systems is described by 1D tight-binding models with quasiperiodic modulations. There is an
abundance of quasiperiodic modulations, among which the canonical types are the Harper model (also known as the Aubry-Andr\'{e}
model)~\cite{Harper:1955,AA} and the Fibonacci quasicrystal (QC)~\cite{Levine84}. The quasiperiodicity of the Harper model enters
in the form of a cosine modulation incommensurate with lattice spacing, whereas the Fibonacci QC has two discrete values that
appear interchangeably according to the Fibonacci sequence. Moreover, the quasiperiodicity may appear in on-site terms
(diagonal), in hopping terms (off diagonal), or in both (generalized). Each of these models describes different physical
phenomena. Indeed, the Harper and the Fibonacci modulations have different localization phase diagrams, depending also on their
appearance as diagonal or off-diagonal terms (see e.g., Refs.~\cite{Kohmoto:1983,Ostlund:1983,Han:1994,Hiramoto:1992}). Notably,
several attempts were made to gather these models under some general framework~\cite{Hiramoto:1989b,Hiramoto:1992,Naumis:2008},
but with only partial success.

These 1D quasiperiodic models play a nontrivial role in the rapidly growing field of topological phases of
matter~\cite{us,citing_us1,citing_us2}. This new paradigm classifies gapped systems such as band insulators and
superconductors~\cite{RMP_TI,RMP_TI2}. Each gap in these systems is attributed an index that characterizes topological properties
of the wave functions in the bands below this gap. By definition, two gapped systems belong to the same topological phase if they
can be deformed continuously from one into the other without closing the energy gap. Conversely, while deforming a system with a
given topological index to a system with another index the bands invert and the bulk gap closes, i.e.~a quantum phase transition
occurs.

The spectra of the aforementioned 1D quasiperiodic models are gapped and, hence, appropriate for topological classification.
Apparently, in the absence of any symmetry all 1D systems are topologically trivial~\cite{Avron1}; namely, all their topological
indices are zero, and therefore the hopping terms in such systems can be continuously turned-off (the atomic limit). Conversely,
under the same conditions 2D systems have nontrivial topological phases, which are characterized by an integer index -- the Chern
number. If a Hamiltonian of a 1D system depends on a periodic parameter, then this parameter can be considered as an additional
dimension. Taking into account all the possible values of this parameter, the system becomes effectively 2D, and may have a
nontrivial Chern number~\cite{Thouless_pump}. This is seen during the evaluation of the Chern number, which requires integration
of the Berry curvature over this parameter. It was shown in Ref.~\cite{us} that for QCs, a translation of the quasiperiodic
modulation may be considered as such a parameter. Remarkably, for QCs the Berry curvature is invariant to this translation,
making the integration over it redundant. Therefore, it was concluded that 1D quasiperiodic systems can be associated with Chern
numbers, making the 1D QCs topologically classified.

Here we use this approach to prove that \emph{all} the aforementioned 1D models are topologically equivalent whenever they have
the same modulation frequency. Hence, there is no quantum phase transition when deforming between these different models. To
prove this equivalence, we extend each one of the 1D models to an ``ancestor'' 2D model. We find that all the resulting models
are variants of the 2D integer quantum Hall effect on a lattice. These 2D models are topologically equivalent and nontrivial.
Therefore, their corresponding 1D descendants are also topologically equivalent and nontrivial. Remarkably, the equivalence holds
between any Fibonacci-like quasicrystal and a Harper model with a corresponding modulation frequency.

One-dimensional tight-binding Hamiltonians with nearest-neighbor hopping and an on-site
potential can be written in the general form
\begin{equation} \label{Eq:H1D}
    H = \sum_n \Big[ \left(t+\lod \Vod_n \right) c_n^\dag c_{n + 1} + h.c.
      + \ld \Vd_n c_n^\dag c_n \Big ],
\end{equation}
where $c_n$ is the single-particle annihilation operator at site $n$, $t$ is some real hopping amplitude, $\Vod$ is some hopping
modulation (off-diagonal term) and $\Vd$ is an on-site potential (diagonal term). The real and positive parameters $\lod$ and
$\ld$ control the strength of the off-diagonal and diagonal modulations, respectively. The quasiperiodicity of the different
models is encoded in the potential modulations, $\Vod$ and $\Vd$. Using this general form, we first show the topological
equivalence between all the Harper models, since their relation to the quantum Hall effect is evident. Then we turn to encompass
also the Fibonacci QCs.

Let us begin with the Harper models, which are governed by the modulation $\VH_n(k) = \cos (2\pi bn + k)$. This modulation is
parametrized by the frequency $b$ and phase $k$. Whenever $b$ is irrational, the modulation is incommensurate with the lattice
and describes a QC. In this case, $k$ resembles a translation of the quasiperiodic modulation. Since $k$ does not affect bulk
properties, it was usually ignored in previous analyses. However, as we shall soon observe, it plays a crucial role in unraveling
the topological behavior of quasiperiodic models~\cite{us}.

\begin{figure}
\centering
\includegraphics[width=\columnwidth]{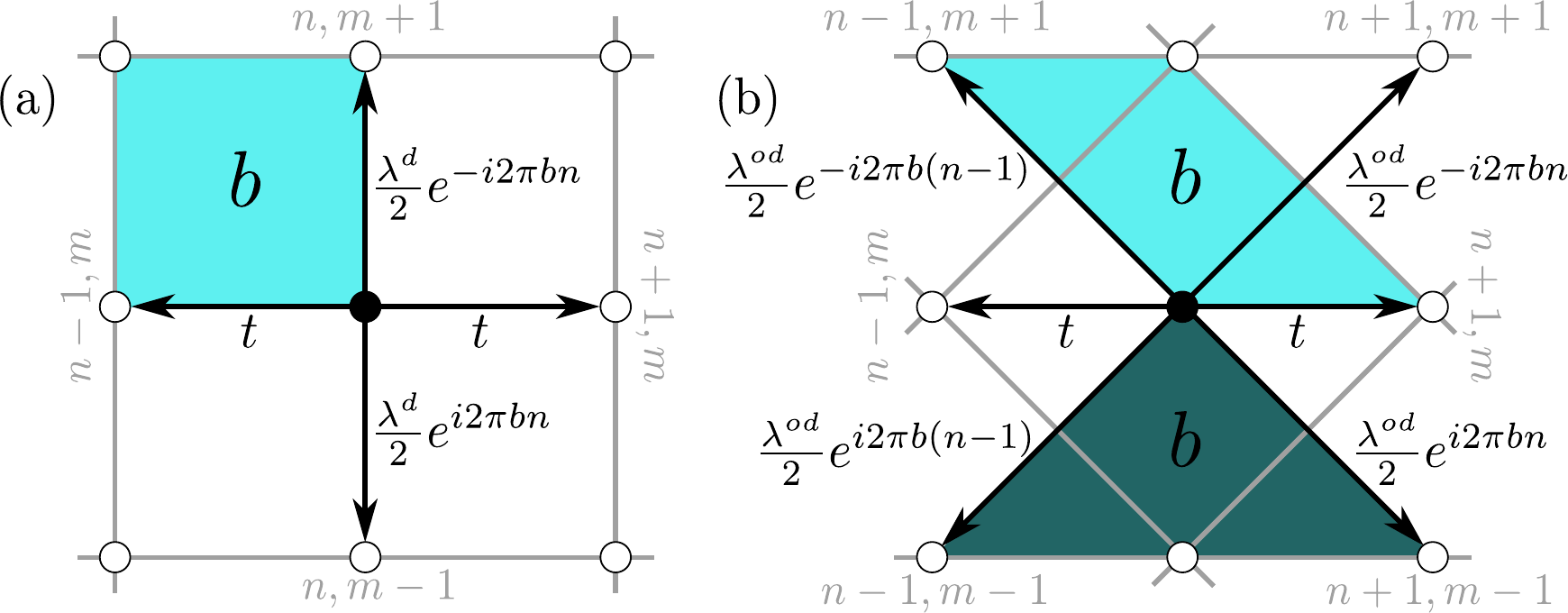}
\caption{\label{fig1} Graphical presentations of 2D Hamiltonians. The electrons hop on a rectangular lattice in the presence of a
perpendicular magnetic field with $b$ flux quantum per unit cell. A hopping amplitude from one vertex to another is denoted by an
arrow. (a) The Hamiltonian $\H\dH$, the ``ancestor'' of the diagonal Harper model. The hopping is to nearest neighbors, and each
rectangular plaquette (marked in cyan) is pierced by $b$ flux quantum. (b) The 2D ancestor Hamiltonian of the off-diagonal Harper
model, $\H\oH$. The hopping is to nearest neighbors in the $n$ direction, and also to next-nearest neighbors. Here, there are two
types of plaquettes, a triangle (dark green) and a parallelogram (cyan), both pierced by $b$ flux quantum. The magnetic
translation group is the same in both Hamiltonians, implying that they are topologically equivalent. }
\end{figure}

The \textit{diagonal Harper model}~\cite{Harper:1955} is defined by setting $\lod=0$, $\ld\neq 0$, and $\Vd_n = \VH_n(k)$ in
Eq.~\eqref{Eq:H1D}. The corresponding Hamiltonian describes uniform hopping and a modulated on-site potential. For any given $k$,
this Hamiltonian can be viewed as the $k$th Fourier component of some ancestor 2D Hamiltonian. From this viewpoint, $k$ is a
second degree of freedom, hence we define the operator $c_{n,k}$ that obeys the commutation relation $\{
c_{n,k},c_{n^\prime,k^\prime}^\dag \} = \delta_{n,n^\prime}\delta_{k,k^\prime}$. We can now define a 2D Hamiltonian $\H =
\int_0^{2\pi} (dk/2\pi) H(k)$, where in $H(k)$ we replace the operators $c_n$ with $c_{n,k}$. Note that, in the following, $H$
($\H$) denotes 1D (2D) Hamiltonians. Defining the Fourier transform $c_{n,k}= \sum_m e^{-ikm} c_{n,m}$, we obtain the 2D ancestor
Hamiltonian of the diagonal Harper model
\begin{equation} \label{ed:HdH_2D}
    \resizebox{.9\hsize}{!}{$\displaystyle\H\dH = \sum_{n,m} \Big[ t c_{n,m}^\dag c_{n + 1,m}
    + \frac{\ld}{2} e^{i2\pi bn} c_{n,m}^\dag c_{n,m + 1} + h.c. \Big].$} 
\end{equation}
This Hamiltonian describes electrons hopping on a 2D rectangular lattice in the presence of a uniform perpendicular magnetic
field with $b$ flux quantum per unit cell~\cite{Harper:1955,Azbel,Hofstadter}, as illustrated in Fig.~\ref{fig1}(a). Note that,
in $\H\dH$ the magnetic field appears in Landau gauge.

In the absence of a magnetic field, the Hamiltonian commutes with the group of translations. Since the magnetic field breaks this
symmetry, the notion of the magnetic translation group was introduced~\cite{Zak}. The magnetic translation group is generated by
the operators $T_{\hat{m}}$ and $T_{\hat{n}}$, where $T_{\hat{m}} c_{n,m} T_{\hat{m}}^{-1} = c_{n,m+1}$ and $T_{\hat{n}} c_{n,m}
T_{\hat{n}}^{-1} = e^{-i2\pi b m} c_{n+1,m}$. These operators commute with the Hamiltonian but not with each other. However, for
a rational flux, $b=p/q$, the operator $T_{q\hat{n}} = (T_{\hat{n}})^q$ commutes with $T_{\hat{m}}$. Therefore it is possible to
diagonalize simultaneously $\H\dH$, $T_{q\hat{n}}$, and $T_{\hat{m}}$. The spectrum in this case is composed of $q$
bands~\cite{Hofstadter}. In a seminal paper by Thouless \etal~\cite{TKNN}, it was shown that each gap in the spectrum of this
model is associated with a quantized and nontrivial Chern number (Hall conductance). Later on, it was shown by Dana
\etal~\cite{Avron:1985} that the nontriviality of the Chern numbers stems from the symmetry of this model with respect to the
magnetic translation group. Choosing consistent boundary conditions~\cite{Avron:1985}, the Chern number $\nu_r$ that is
associated with a gap number $r=1,\ldots,(q-1)$ abides the Diophantine equation $r = \nu_r q + t_r p$, where $\nu_r$ and $t_r$
are integers, and $0 < |\nu_r| < q/2$ \cite{TKNN}.

An irrational $b$ can be approached by taking an appropriate rational limit with $p,q \rightarrow \infty$. In this limit the
spectrum becomes fractal~\cite{Hofstadter}. Nevertheless, even for an arbitrarily large $q$, the system abides the aforementioned
Diophantine equation, and the gaps remain associated with nontrivial Chern numbers. Formally, the evaluation of the Chern numbers
requires integration of the Berry curvature over $k$~\cite{TKNN}. Hence, the quantized Chern numbers characterize only the 2D
ancestor model, rather than its 1D descendant model. However, as shown in Ref.~\cite{us}, for a QC, i.e., for an irrational $b$,
the Berry curvature is independent of $k$ and there is no need for such an integration. Therefore, the 1D models can be
associated with the same quantized topological indices.

This simple model demonstrates that the method to extract the topological indices of a 1D QC is to extend it to 2D using the
above procedure and find the magnetic translation group of the ancestor Hamiltonian. This yields a Diophantine equation and,
thus, the Chern number of each gap.

Having performed this method for the diagonal Harper model, we turn, now, to the \textit{off-diagonal Harper model}. In this
model the hopping is modulated and the on-site potential vanishes. It is defined by setting $\lod \neq 0$, $\ld = 0$, and $\Vod_n
= \VH_n(k)$ in Eq.~\eqref{Eq:H1D}. Constructing its 2D ancestor model, we obtain
\begin{align} \label{ed:HodH_2D}
    \H\oH =& \sum_{n,m} \Big[ t\, c_{n,m}^\dag c_{n + 1,m}
                + \frac{\lod}{2} \Big( e^{i2\pi bn} c_{n,m}^\dag c_{n+1,m+1} \nonumber \\
    & \qquad + e^{-i2\pi bn} c_{n,m}^\dag c_{n+1,m-1} \Big) + h.c. \Big].
\end{align}
This Hamiltonian describes electrons hopping on a rectangular lattice with nearest-neighbors hopping only in the $n$ direction,
and also next-nearest-neighbors hopping. Here too a perpendicular magnetic field is present with $b$ flux quanta per unit cell
and per plaquette, as illustrated in Fig.~\ref{fig1}(b). The corresponding magnetic translation group is the same as in the
diagonal case. Hence, this model abides the same Diophantine equation, which characterizes its gaps. Therefore, for a given $b$,
the 2D ancestor Hamiltonians of the diagonal and off-diagonal Harper models have in fact the same number of gaps and the same
distribution of Chern numbers, making them topologically equivalent.

The diagonal and off-diagonal Harper models are incorporated in the \textit{generalized Harper model}~\cite{Han:1994}, where in
Eq.~\eqref{Eq:H1D} we take $\lod \neq 0, \ld \neq 0$, $\Vd_n = \VH_n(k)$, and $\Vod_n = \VH_n(k + \pi b)$. Now both the hopping
terms and the on-site potential are cosine modulated. Its corresponding 2D model has both nearest-neighbor and
next-nearest-neighbor hopping. The magnetic flux per unit cell is still $b$, and the magnetic translation group remains the same
as well. Hence, the Diophantine equation is also the same, independent of relative modulation strengths $\ld$ and $\lod$. We can
therefore conclude that, for a given $b$, as the ratio $\lod/\ld$ is changed all the energy gaps remain open and thus keep their
Chern numbers fixed~\cite{Han_inversion}. This means that the generalized Harper model provides a way to continuously deform the
diagonal Harper model into the off-diagonal one, and vice versa, without experiencing a quantum phase transition.

After showing the topological equivalence between the Harper models, we now address the Fibonacci QC. This QC is governed by the
modulation $\VF_n = 2( \lfloor (n+2)/\tau \rfloor - \lfloor (n+1)/\tau \rfloor ) - 1 = \pm1$, where $\tau = (1 + \sqrt{5})/2$ is
the golden ratio and $\lfloor x \rfloor$ is the floor function. Similar to the Harper modulation, the $\VF$ modulation can be
employed as a \textit{diagonal Fibonacci QC}, with $\lod = 0$ and $\Vd_n = \VF_n$. Alternatively, it can be employed as an
\textit{off-diagonal Fibonacci QC}, with $\ld = 0$ and $\Vod_n = \VF_n$. Due to the discontinuous nature of these QCs, they have
no apparent ancestor 2D models. This seemingly prevents the extraction of their topological indices.

We can overcome this barrier by constructing a modulation that continuously deforms the Fibonacci into a Harper modulation.
Consider the function $f(x) = 2( \lfloor x + a \rfloor - \lfloor x \rfloor ) - 1$, where $0 < a < 1$. The function $h(x) =
\cos(2\pi x + a\pi) - \cos(a\pi)$ has the same sign as $f(x)$ for any $x$. Therefore $g(x;\beta) = \tanh[ \beta h(x)
]/\tanh[\beta]$ is a continuation between the smooth $h(x)=g(x;\beta \rightarrow 0)$ and the steplike $f(x)=g(x;\beta \rightarrow
\infty)$.

\begin{figure}
\centering
\includegraphics[width=\columnwidth]{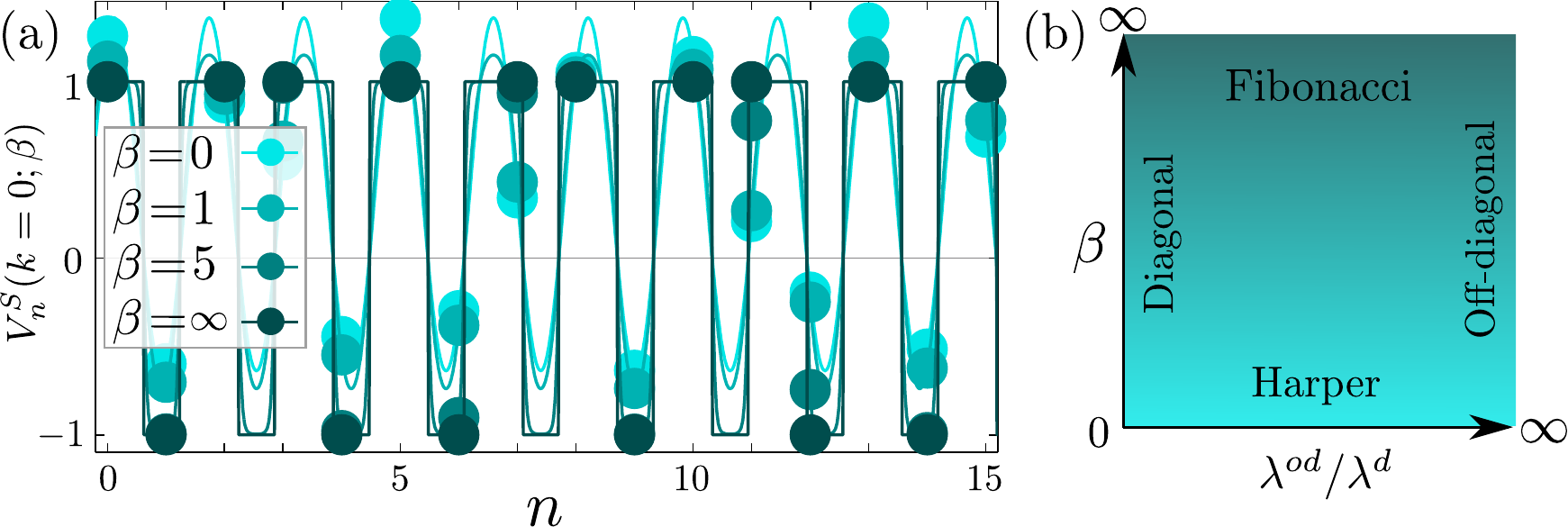}
\caption{\label{fig2} (a) The smooth continuation between the Harper and the Fibonacci modulations $\VS_n$
[cf.~Eq.~\eqref{smooth}] for $b=2/(1 + \sqrt{5})$, $k=0$, and several values of the smoothing parameter $\beta$. For $\beta
\rightarrow 0$, $\VS$ resembles the Harper modulation $\VH$, whereas for $\beta \rightarrow \infty$, it approaches the Fibonacci
modulation $\VF$. Varying $\beta$ has no influence on the distribution of the Chern numbers, meaning that the Harper model and
the Fibonacci QC are topologically equivalent. (b) Illustration of the space of quasiperiodic systems spanned by the smoothing
parameter $\beta$ and the diagonal-off-diagonal ratio $\lod/\ld$, when substituted in the general Hamiltonian $H^\text{S}$
[cf.~Eq.~\eqref{Eq:HS_1D}]. All the systems in this space are topologically equivalent, with the same number of gaps and the same
Chern numbers. }
\end{figure}

Accordingly, we define the smooth modulation
\begin{align} \label{smooth}
    \VS_n(k;\beta) = \frac{\tanh \{ \beta [ \cos (2\pi bn + k) -
    \cos \pi b ]\}} {\tanh \beta} \, .
\end{align}
It can be seen that, in the limit of small $\beta$, this smooth modulation becomes the Harper modulation, up to a constant shift,
$\VS_n(k;\beta \rightarrow 0) = \VH_n(k) - \cos \pi b$. In the opposite limit, it approaches the Fibonacci modulation, $\VS_n(k =
3\pi b; \beta \rightarrow \infty) = \VF_n$ for $b = 1/\tau$, as depicted in Fig.~\ref{fig2}(a). Similar to the generalized Harper
model, we now define a generalized smooth model
\begin{align} \label{Eq:HS_1D}
    H^\text{S}(k;\beta) = \sum_n \{ & [t+\lod \VS_n(k + 4\pi b;\beta)]
    c_n^\dag c_{n + 1} + h.c. \nonumber \\
      & + \ld \VS_n(k + 3\pi b;\beta) c_n^\dag c_n \},
\end{align}
where $b=1/\tau$. This model is a continuous deformation between the generalized Harper model and a generalized Fibonacci QC,
with $\beta$ being the deformation parameter.

Now we are able to construct a corresponding 2D model that includes also the Fibonacci QC. Recall that, in the diagonal Harper
model, the modulation of $\cos (2\pi bn+k)$ resulted in hopping one site in the $m$ direction, accompanied with a phase factor of
$e^{\pm i2\pi bn}$ [see Eq.~(\ref{ed:HdH_2D})]. One can think, however, of a more general modulation $F[\cos (2\pi bn+k)]$, where
$F$ is some analytic function. In such cases, $F$ can be expanded as a Taylor series in powers of $\cos (2\pi bn+k)$. In turn,
the series can be rewritten as a series in powers of $e^{\pm i(2\pi bn+k)}$. The $l$th component of this series will appear in
the 2D ancestor model as a term of hopping $l$ sites in the $m$ direction. The amplitude of this hopping term would be $F_l
e^{i2\pi bnl}$, where $F_l$ is the Fourier transform of $F[\cos (k)]$. Similarly, an off-diagonal modulation will result in terms
that incorporate hopping to the nearest-neighbor in the $n$ direction with longer-range hopping in the $m$ direction. Therefore,
the 2D ancestor Hamiltonian of $H^\text{S}(k;\beta)$ is
\begin{align} \label{Eq:HS_2D}
    \H^\text{S}(\beta) = &\sum_{n,m} \Big[ t c_{n,m}^\dag c_{n + 1,m} \\
    & + \sum_{l=0}^\infty \vSl(\beta) \Big( \lod e^{i\pi b(2n+4)l} c_{n,m}^\dag c_{n+1,m+l} \nonumber\\
    & \qquad \qquad + \lod e^{-i\pi b(2n+4)l} c_{n,m}^\dag c_{n+1,m-l} \nonumber \\
    & \qquad \qquad + \ld  e^{i\pi b(2n+3)l} c_{n,m}^\dag c_{n,m+l} \Big) + h.c. \Big], \nonumber
\end{align}
where $\vSl(\beta) = \int_0^{2\pi} (dk/2\pi) e^{ilk} \VS_{n=0}(k;\beta)$. It can be shown that~\cite{supmat}
\begin{equation} \label{Eq:vl_full}
\resizebox{.9\hsize}{!}{$\displaystyle
\vSl(\beta)  = \left \{ \begin{array}{lc} %
l \in 2\mathbb{Z}   & (-1)^{\frac{l}{2}-1} \frac{4}{\beta \tanh\beta}
    \sum_{j=0}^\infty \text{Im} \left( \frac{z_j^{\phantom{j}l+1}}{1+z_j^{\phantom{n}2}} \right) \\
l \in 2\mathbb{Z}+1 & (-1)^{\frac{l-1}{2}} \frac{4}{\beta \tanh\beta}
    \sum_{j=0}^\infty \text{Re} \left( \frac{z_j^{\phantom{j}l+1}}{1+z_j^{\phantom{n}2}} \right),
\end{array} \right. $}
\end{equation}
where $z_j(\beta) = \sqrt{1 + (w_j - i\alpha)^2} - (w_j - i\alpha)$ with \mbox{$w_j = (j + 1/2)\pi/\beta$} and $\alpha = \cos(\pi
b)$. The physical meaning of $\vSl$ is better understood by looking at the limits of large $\beta$ and small $\beta$. In the
Harper limit of $\beta \ll 1$~\cite{supmat},
\begin{equation}
\resizebox{.9\hsize}{!}{$\displaystyle\vSl(\beta \ll 1)  \approx  \left\{ \begin{array}{lc}
    l \in 2\mathbb{Z}   & -\alpha \frac{8}{\pi^2} ( |l| + 1 ) u_l \left( \frac{i\beta}{\pi} \right)^{|l|} \\
    l \in 2\mathbb{Z}+1 & \frac{4}{\pi^2} u_l \left( \frac{i\beta}{\pi} \right)^{|l|-1},
    \end{array} \right.$}
\end{equation}
where $u_l  = 1 + \left( \delta_{l,0} + \delta_{l,\pm1} \right) \left( \pi^2/8 - 1\right)$. We can see that, in this limit, the
hopping in the $m$ direction decays exponentially with the distance $l$. In the extreme limit of $\beta = 0$, only the terms with
$l=0,\pm1$ survive and $\H^\text{S}$ becomes the 2D ancestor of the generalized Harper model (up to a constant shift of the
energy). In the opposite limit of $\beta \gg 1$, i.e., the Fibonacci limit~\cite{supmat},
\begin{equation}
    \resizebox{.9\hsize}{!}{$\displaystyle\vSl(\beta \gg 1)  \approx (-1)^{\lfloor b \rfloor} \left( \frac{2}{\pi l}
    \sin(l \pi b) + \delta_{l,0} \left( 1 + 2\lfloor b \rfloor \right) \right).$}
\end{equation}
Now the hopping in the $m$ direction is no longer local, but decays as $1/l$.

Regardless of the exact value of $\beta$, $\H^\text{S}$ describes electrons hopping on a rectangular lattice. Moreover, since the
amplitudes of the hopping in the $m$ directions are accompanied by the phases $e^{\pm i2\pi bnl}$, in this model also a magnetic
field is present with $b$ flux quanta per unit cell. Remarkably, despite the varying hopping behavior, the magnetic translation
group remains the same for all values of $\beta$. Therefore, the Diophantine equation is also the same for all $\beta$.
Consequently, while varying $\beta$ from zero to infinity the gap structure and its corresponding nontrivial Chern numbers are
unchanged. Since $\beta$ turns the Harper model into the Fibonacci QC, it implies that they are topologically equivalent. Note
that this is true also for any value of $\lod/\ld$.

The real and positive parameters, $\beta$ and $\lod/\ld$, span the space of models that contain the diagonal and off-diagonal,
Harper and Fibonacci models, as illustrated in Fig.~\ref{fig2}(b). We can therefore conclude that all the 1D models $H^\text{S}$
in this space are topologically equivalent and nontrivial. Moreover, the same holds true for \emph{any} irrational $b$. Here,
taking $\beta$ to infinity results in a Fibonacci-like QC with $\tau=1/b$ as the modulation frequency of $\VF$. Note that any
Fibonacci-like QC can be obtained via the cut-and-project method~\cite{QC_Senechal}, by taking $\tau = 1+1/\cot\theta$ with
$\theta$ the angle of the projection line. This means that any Fibonacci-like QC with a given $\tau$ is topologically equivalent
to a Harper model with $b=1/\tau$, and both are topologically nontrivial. For a rational $b$, i.e., for periodic systems, the 2D
ancestor Hamiltonians are equivalent, but the implications to the 1D models are more subtle~\cite{us}. Nevertheless, no bulk gaps
will close when deforming between them.

The physical manifestation of the topological nontriviality of a QC is easily seen in the emergence of boundary states that
traverse the energy gaps as a function of the translation parameter $k$. The Chern number of each gap is equal to the number of
boundary states that traverse the gap. The topological equivalence of the aforementioned 1D models implies that the number of
traversing boundary states in a given gap is constant when $\beta$ and $\lod/\ld$ are changed. This should be contrasted to the
localization properties of the bulk states, which vary considerably during such a change~\cite{Han:1994}.

The topological nontriviality of the diagonal and off-diagonal Harper models has been recently demonstrated experimentally via
boundary phenomena~\cite{us}. It would be intriguing to test this for the Fibonacci QC as well, where we expect gap-traversing
boundary modes to appear. Our prediction is supported by the fact that the existence of subgap boundary states in the Fibonacci
QC was noticed~\cite{Fibonacci_theory}, measured~\cite{Fibonacci_experiment1,Fibonacci_experiment2}, and analyzed to be
quantitatively similar to those of the Harper model~\cite{Molina}.

To summarize, we developed a 1D model that ranges smoothly from the Harper model to a Fibonacci or Fibonacci-like QC and from
diagonal to off-diagonal modulations, as a function of control parameters. Using the fact that dimensional extension of QCs from
one to two dimensions reveals their topological character, we extended this model to 2D. We found that in 2D the hopping behavior
changes significantly with the control parameters. Nevertheless, the magnetic translation group is unaffected. This implies that
the same nontrivial Chern numbers remain for all values of the control parameters. Therefore we conclude that all these 1D models
are topologically equivalent and nontrivial. It would be interesting to follow the same process for other QCs, which may exhibit
novel types of topological phases. For example, 2D and 3D QCs which are obtained via cut-and-project
methods~\cite{QC_Senechal,BenAbraham:2007}, may have topological characteristics of 4D and 6D systems, respectively.

We thank A.~Stern, Z.~Ringel, Y.~Lahini, Y.~Avron, I.~Gruzberg, and S.~Jitomiskaya for fruitful discussions. We thank the Minerva
Foundation, the US-Israel Binational Science Foundation, and the IMOS Israel-Korea grant, for financial support.



\newpage
\begin{center}
\textbf{\large SUPPLEMENTARY MATERIAL}
\end{center}
\vspace{10mm}

In the main text we defined a quasiperiodic modulation, $\VS_n(k;\beta)$ [cf.~Eq.~(4) of the main text], which smoothly deforms
the Harper modulation into the Fibonacci modulation as $\beta$ is changed from zero to infinity. Placing it in the generalized
smooth model [cf.~Eq.~(5) of the main text] and constructing an ancestor 2D Hamiltonian requires the evaluation of the following
integral
\begin{align} \label{Eq:Il}
    \mathcal{I}_l = \int_0^{2\pi} \frac{dk}{2\pi} e^{ilk} \VS_n(k + 3\pi b;\beta).
\end{align}
Recall that $\VS_n(k + 3\pi b;\beta)=\tanh  \{ \beta [ \cos(\gamma_n + k) - \alpha ]\} / \tanh \beta$, where $\gamma_n=2\pi bn +
3\pi b$ and $\alpha=\cos \pi b$. Shifting the integration variable $k \rightarrow  k - \gamma_n$ and taking into account the
$2\pi$ periodicity of the integrand, we obtain
\begin{align} \label{Eq:Il_vl}
    \mathcal{I}_l &= \frac{e^{-il\gamma_n}}{\tanh\beta}\int_0^{2\pi}\frac{dk}{2\pi}e^{ilk}\tanh \{ \beta [ \cos (k) -\alpha ] \} \nonumber\\
                  &\equiv e^{-il\gamma_n}\vSl(\beta)\, .
\end{align}
Since the integrand is real and symmetric, we know that $v^\text{S}_{-l} = \vSl$. Therefore it is sufficient to evaluate $\vSl$
for $l \geq 0$. Defining $z=e^{ik}$,
\begin{equation} \label{Eq:vSl}
    \resizebox{.85\hsize}{!}{$\displaystyle \vSl = \frac{1}{\tanh \beta} \oint_{\text{unit}} \frac{dz}{2\pi i} z^{l-1}
    \tanh \left[ \frac{\beta}{2} ( z + z^{-1} - 2\alpha ) \right],$}
\end{equation}
where the integration is performed over the unit circle in the complex plane.

For $l=0$, there is a pole at $z=0$. Below we show that this pole has zero contribution. For any $l \geq 0$, the integrand has
simple poles when
\begin{align} \label{Eq:z_wq}
    z + z^{-1} = 2i(w_j - i\alpha),
\end{align}
with $w_j = (j+1/2)\pi/\beta$ and $j$ is an integer. Specifically, these poles are
\begin{align} \label{Eq:zj}
    z_j^{\pm} & =i\left[ (w_j-i\alpha)\pm\sqrt{(w_j-i\alpha)^2+1} \right].
\end{align}
The poles that contribute to the integral are only those that are within the unit circle. Below, we prove that
$\left|z_j^{+}\right|<1$ for $j \leq -1$, and that $\left|z_j^{-}\right|<1$ for $j \geq 0$. Therefore, using the fact that
$z_{-j}^{+}=\left(z_{j-1}^{-}\right)^*$, the relevant poles are $z_j^{-}$ and $\left(z_{j}^{-}\right)^*$ for $j\geq 0$.

Denoting $z_j^{-}=-iz_j$, where $z_j=\sqrt{(w_j-i\alpha)^2+1} - (w_j-i\alpha)$, the residue of the integrand at $z_j^{-}$ is
given by
\begin{align} \label{Eq:Res}
    \text{Res} \left\{ z^{l-1} \tanh \left[ \frac{\beta}{2} ( z + z^{-1} - 2\alpha  ) \right]; z_j^{-} \right\} \nonumber \\
    \qquad = (-i)^{l-1} \frac{2}{\beta} \frac{ z_j^{\phantom{*}l+1} }{1 + z_j^{\phantom{*}2}}.
\end{align}
We therefore conclude that
\begin{align}
    \vSl = & \frac{2i^{l-1}}{\beta \tanh\beta} \sum_{j=0}^\infty
    \left[ (-1)^{l-1} \frac{ z_j^{\phantom{*}l+1} }{1 + z_j^{\phantom{*}2}}
    + \frac{ (z_j^*)^{l+1} }{1 + (z_j^*)^2} \right],
    \label{general}
\end{align}
which is equivalent to Eq.~(7) in the main text.

Having the exact serial expression for $\vSl(\beta)$, we turn to approximate it in the limits of small and large $\beta$, i.e.
the Harper and Fibonacci limits, respectively. This way, we obtain simpler expressions, which provide better understanding of the
physical effect of $\beta$ on the 2D ancestor Hamiltonian.

\vspace{5mm}
\noindent \textbf{Small $\beta$ limit (Harper limit)} \\
In the limit of $\beta \ll 1$, by definition, $w_j \gg 1$ for every $j \geq 0$. Therefore,
\begin{align}
    z_j(\beta \ll 1) = \frac{1}{2w_j} \left( 1 + i\frac{\alpha}{w_j} \right) +
    O(w_j^{\phantom{*}-3}),
\end{align}
and
\begin{align}
    \frac{ z_j^{\phantom{*}l+1} }{1 + z_j^{\phantom{*}2}} \approx
    \frac{1}{(2w_j)^{l+1}} \left[ 1 + i(l+1)\frac{\alpha}{w_j} \right].
\end{align}
Plugging this into Eq.~\eqref{general}, we consider the cases of even and odd values of $l$.

For odd values of $l$,  we obtain
\begin{equation}
    \resizebox{.85\hsize}{!}{$\displaystyle v^\text{S}_{l \in 2\mathbb{Z}+1}(\beta \ll 1) \approx \frac{1}{\pi^2} \left( \frac{i
    \beta}{2\pi} \right)^{l-1} \sum_{j=0}^\infty \frac{1}{(j+1/2)^{l+1}}.$}
\end{equation}
Notice that $\sum_{j=0}^\infty (j+1/2)^{-l-1} = (2^{l+1} - 1) \zeta(l+1)$, where $\zeta(x)$ is the Riemann zeta function. Using
the facts that $\zeta(2) = \pi^2/6$, and that for $l \geq 3$, $\zeta(l+1) \approx 1$ and $2^{l+1} \gg 1$, we can write
\begin{align} \label{Eq:small_odd}
    v^\text{S}_{l \in 2\mathbb{Z}+1}(\beta \ll 1) \approx \left\{  \begin{array}{lc}
        |l|=1 & 1/2 \\
        |l|=3,5,\ldots & \frac{4}{\pi^2} \left( \frac{i\beta}{\pi} \right)^{|l|-1}
    \end{array} \right. .
\end{align}

For even values of $l$, we obtain
\begin{equation}
    \resizebox{.85\hsize}{!}{$\displaystyle v^\text{S}_{l \in 2\mathbb{Z}}(\beta \ll 1) \approx
    -\alpha (l+1) \frac{2}{\pi^2} \left( \frac{i
    \beta}{2\pi} \right)^l \sum_{j=0}^\infty \frac{1}{(j+1/2)^{l+2}}.$}
\end{equation}
Similarly, $\sum_{j=0}^\infty (j+1/2)^{-l-2}=(2^{l+2} - 1) \zeta(l+2)$, resulting in
\begin{align} \label{Eq:small_even}
    \resizebox{.85\hsize}{!}{$\displaystyle v^\text{S}_{l \in 2\mathbb{Z}}(\beta \ll 1) \approx \left\{  \begin{array}{lc}
        l=0 & -\alpha \\
        |l|=2,4,\ldots & -\alpha \frac{8}{\pi^2} (l+1) \left( \frac{i\beta}{\pi}
        \right)^{|l|}
    \end{array} \right. .$}
\end{align}
The results of Eq.~(\ref{Eq:small_odd}) and Eq.~(\ref{Eq:small_even}) are summarized in Eq.~(8) of the main text.

\vspace{5mm}
\noindent \textbf{Large $\beta$ limit (Fibonacci limit)} \\
In the opposite limit of $\beta \gg 1$, the poles $w_j$ becomes dense. Therefore we can approximate the sum $\sum_{j=0}^\infty$
with the integral $(\beta/\pi)\int_{\pi/2\beta}^\infty \d w_j$. Additionally, using the relation $1+z_j^{\phantom{*}2} = 2z_j
\left[ 1 + (w_j - i\alpha)^2 \right]^{1/2}$, we find that $\d z_j/ \d w_j = -2z_j^{\phantom{*}2} ( 1+z_j^{\phantom{*}2} )^{-1}$.
Hence, for $l > 0$,
\begin{align} \label{Eq:sum_int}
    \sum_{j=0}^\infty \frac{z_j^{\phantom{*}l+1}}{1 + z_j^{\phantom{*}2}} & \approx
    -\frac{\beta}{2\pi} \int_{\pi/2\beta}^\infty \d w_j z_j^{\phantom{*}l-1} \frac{\d
    z_j}{\d w_j} \nonumber \\
    &= \frac{\beta}{2\pi l} z_j^{\phantom{*}l}(w_j = \pi/2\beta).
\end{align}

In the limit of $\beta \rightarrow \infty$, we observe that $w_j \ll 1$ for any finite $l$. Since $|\alpha| < 1$, we obtain
\begin{align}
    z_j(\beta \rightarrow \infty) = & \left( \sqrt{1 + \alpha^2} + i\alpha \right) \left( 1 - \frac{w_j}{\sqrt{1 + \alpha^2}} \right)\nonumber\\
    & +    O(w_j^{\phantom{*}2}).
\end{align}
Rewriting $\sqrt{1 + \alpha^2} + i\alpha = e^{i\arcsin{\alpha}}$, we have
\begin{align} \label{Eq:sum_res}
    \sum_{j=0}^\infty \frac{z_j^{\phantom{*}l+1}}{1 + z_j^{\phantom{*}2}}
    \xrightarrow{\beta \rightarrow \infty} \frac{\beta}{2\pi l} e^{il\arcsin{\alpha}}.
\end{align}

For $l=0$,
\begin{align} \label{Eq:sum_int_0}
    \sum_{j=0}^\infty \frac{z_j}{1 + z_j^{\phantom{*}2}} & \approx
    -\frac{\beta}{2\pi} \int_{\pi/2\beta}^\infty \frac{\d w_j}{\sqrt{1 + (w_j - i\alpha)^2}} \nonumber \\
    &= \frac{\beta}{2\pi} [ \textrm{asinh}(w)|_{w \rightarrow \infty} - \textrm{asinh}(\pi/2\beta - i\alpha) ]  \nonumber \\
    &\approx \frac{\beta}{2\pi} [ \textrm{asinh}(w)|_{w \rightarrow \infty} + i\arcsin\alpha ].
\end{align}

Therefore, according to Eq.~\eqref{general}, we obtain
\begin{align} \label{Eq:large}
    \vSl(\beta \rightarrow \infty) = \frac{2}{\pi l} \sin( l \arccos \alpha ) - \delta_{l,0},
\end{align}
where we used the facts that $\arcsin \alpha = \pi/2 - \arccos \alpha$, and that $\sin(lx)/l = x$ for $l=0$. Recalling that
$\alpha=\cos \pi b$, we obtain Eq.~(9) of the main text.

\vspace{5mm}
\noindent \textbf{The pole at $z=0$} \\
We have mentioned above that the integrand in Eq.~(\ref{Eq:vSl}) has a pole at $z=0$ for $l=0$, and claimed that it has null
contribution to the integral. In order to prove this, we consider the integral
\begin{equation} \label{Eq:v_epsilon}
    v^\epsilon = \frac{1}{\tanh \beta} \oint_{C_\epsilon} \frac{d \bar{z}}{2\pi i} \bar{z}^*
    \tanh \left[ \frac{\beta}{2} ( \bar{z} + \bar{z}^* - 2\alpha ) \right],
\end{equation}
where $C_\epsilon$ is a circle in the complex plane defined by $|\bar{z}| = \epsilon$. By denoting $\bar{z} = \epsilon z$, we
obtain
\begin{align} \label{Eq:v_epsilon_z}
    v^\epsilon &= \frac{\epsilon^2}{\tanh \beta} \oint_{\text{unit}} \frac{dz}{2\pi i} z^{-1}
    \tanh \left[ \frac{\beta}{2} ( \epsilon z + \epsilon z^{-1} - 2\alpha ) \right] \nonumber \\
    &= \frac{\epsilon^2}{\tanh \beta} \int_0^{2\pi} \frac{dk}{2\pi} \tanh \{ \beta [ \epsilon \cos (k) -\alpha ] \}.
\end{align}
Recalling that $|\alpha| < 1$, we can bound $|v^\epsilon| \leq \epsilon^2$, meaning that $v^\epsilon$ vanishes when $\epsilon
\rightarrow 0$. In this limit the only pole of the integrand that resides within $C_\epsilon$ is $z=0$. We can therefore conclude
that the contribution of this pole to the integral in Eq.~(\ref{Eq:vSl}) is zero.

\vspace{5mm}
\noindent \textbf{Poles within the unit circle} \\
In this last part we prove that the poles $z_j^{\pm}$ that reside within the unit circle are those with $j \le -1$ for $z_j^+$
and with $j \ge 0$ for $z_j^-$.

We do so by proving that $|z_j^{\pm}| = 1$ iff $w_j = 0$. Note that $w_j \neq 0$ for any integer $j$, but changes its sign from
negative to positive only upon an increment from $j=-1$ to $j=0$. From Eq.~(\ref{Eq:z_wq}) we can deduce that $z_{j \rightarrow
\infty}^+ \rightarrow \infty$ and $z_{j \rightarrow -\infty}^+ \rightarrow 0$, while $z_{j \rightarrow \infty}^- \rightarrow 0$
and $z_{j \rightarrow -\infty}^- \rightarrow \infty$. Therefore, this proves the preposition.

For brevity, we denote $\w = w_j - i\alpha$, and rewrite Eq.~(\ref{Eq:z_wq}) as $z(z-2i\w) = -1$. Multiplying each part of this
equation by its complex conjugate gives
\begin{align}
    |z|^2 \left[ |z|^2 + 4|\w|^2 + 4\text{Im}(z^*\w) \right] = 1.
\end{align}
Requiring $|z| = 1$, we obtain $|\w|^2 = -\text{Im}(z^*\w)$. Since $z^*\w = |\w|^2 z^*/\w^*$, we get $\text{Im}(z/\w) = 1$.

Recalling from Eq.~(\ref{Eq:zj}) that $z = i(\w \pm \sqrt{1 + \w^2})$, we end up with $\text{Re}(\sqrt{1 + \w^{-2}}) = 0$. This
is true only if $\w$ is purely real and $\w^{-2} \leq -1$. By definition this is the case only when $|\alpha| \leq 1$ (which is
consistent) and $w_j = 0$.

\end{document}